\documentclass[reprint,prd,aps,showpacs]{revtex4-1}

\usepackage{graphicx,bm,amssymb,amsmath}
\usepackage{ulem} 

\newcommand{\bX}{{\bf X}}

\def\tensor#1{\protect\@ontopof{#1}{\leftrightarrow}{1.15}\mathord{\box2}}

\newcommand{\kB}{{k_{\rm B}}}
\newcommand{\kb}{{k_{\rm B}}}

\usepackage[usenames]{color}

\definecolor{brown}{rgb}{0.3,0.2,0}

\usepackage{graphicx}

\begin{document}

\title{The statistics of mesoscopic systems and the physical interpretation of extensive and non-extensive entropies}

\author{D.~V.~Anghel$^1$ and A.~S.~Parvan$^{1,2,3}$}
\affiliation{$^{1}$Institutul National de Fizica  si Inginerie  Nucleara--Horia Hulubei, Magurele, Romania}
% 
% % \affiliation{$^{1}$Department of Theoretical Physics, Horia Hulubei National Institute of Physics and Nuclear Engineering, Bucharest-Magurele, Romania}
% 
\affiliation{$^{2}$Bogoliubov Laboratory of Theoretical Physics, Joint Institute for Nuclear Research, Dubna, Russia}
\affiliation{$^{3}$Institute of Applied Physics, Moldova Academy of Sciences, Chisinau, Republic of Moldova}

\begin{abstract}
The  postulates of thermodynamics were originally formulated for macroscopic systems.
They lead to the definition of the entropy, which, for a homogeneous system, is a homogeneous function of order one in the extensive variables and is maximized at equilibrium.
We say that the macroscopic systems are extensive and so it is also the entropy.
For a mesoscopic system, by definition, the size and the  contacts with other systems influence its thermodynamic properties and therefore, if we define an entropy, this cannot be a homogeneous of order one function in the extensive variables.
So, mesoscopic systems and their entropies are non-extensive.
While for macroscopic systems and homogeneous entropies the equilibrium conditions are clearly defined, it is not so clear how the non-extensive entropies should be applied for the calculation of equilibrium properties of mesoscopic systems--for example it is not clear what is the role played by the boundaries and the contacts between the subsystems.
We  propose here a general definition of the entropy in the equilibrium state, which is applicable to both, macroscopic and mesoscopic systems.
This definition still leaves an apparent ambiguity in the  definition of the entropy of a mesoscopic system, but this we recognize as the signature of the anthropomorphic character of the entropy (see Jaynes, Am. J. Phys.~{\bf 33}, 391, 1965).

To exemplify our approach, we analyze four formulas for the entropy (two for extensive and two for non-extensive entropies) and calculate the equilibrium (canonical) distribution of probabilities by two methods for each.
We  show  that these methods, although widely used, are not equivalent and one of them is a consequence of our definition of the entropy of a compound system.
\end{abstract}

\pacs{05.; 05.90.+m; 02.50.-r}

\maketitle

\section{Introduction}~\label{sec_intro}
The concept of entropy, central to thermodynamics, was introduced in the nineteenth century to describe the evolution of systems to equilibrium (see, for example, Ref.~\cite{Tisza:book}, which includes a very nice historical perspective).
The entropy $S$ is a function of state and originally referred to macroscopic systems, consisting of a large number of particles (comparable to the Avogadro's number).
In any reversible process, $S$ remains constant, whereas in irreversible processes it always increases.
The variation of entropy $\Delta S$ is considered a measure of the irreversibility of a process.
Nevertheless, as Jaynes pointed out~\cite{AmJPhys.33.391.1965.Jaynes}, the entropy has an intrinsic anthropomorphic nature and is influenced by the observer.

For a large class of systems (that we call \textit{extensive systems}), the entropy is \textit{additive}: if two systems are in equilibrium, $S_1$ is the entropy of \textit{system}~1 and $S_2$ is the entropy of \textit{system}~2, then the entropy of the total \textit{system}~$(1 + 2)$ is $S_{1+2} = S_1 + S_2$ and it does not depend on the contact (or, implicitly, the absence of contact) between the systems.
This is the case of systems in which the boundaries and the interfaces between subsystems are irrelevant at equilibrium and therefore the system may be regarded as being composed of independent subsystems in equilibrium.
A consequence of this property is the fact that the entropy is a homogeneous function of order 1 in the extensive parameters.
If we denote by $E$ the internal energy of the system, by $\bX \equiv (X_1, X_2, \ldots)$ the set of all the other extensive parameters that define the state of the system, and by $\lambda$ a positive constant, then $S$ is homogeneous of order 1 if $S(\lambda E, \lambda \bX) = \lambda S(E, \bX)$.
The extensive systems are analyzed in the context of the Gibbsian thermostatistics~\cite{Tisza:book,Prigogine:book}, which was  later extended by R\'enyi~\cite{ProcFourthBerkeleySymposium.1.547.1061.Renyi, Renyi:book}.

The \textit{mesoscopic systems} (i.e. systems in which finite size effects play an important role, like nano-systems, nuclei, or systems composed of astronomical objects), which are intensely studied nowadays, are \textit{non-extensive}, i.e. if \textit{system}~1 and \textit{system}~2
% are isolated, then the entropy of the \textit{system}~$(1+2)$ is $S_{1+2}$, different from $S_1 + S_2$.
are in contact and in equilibrium with each other (it does not matter, for the present discussion, what kind of contact exists between the systems) and then we isolate them without changing their extensive parameters $(E_i. \bX_i)$ $(i=1,2)$--assuming we can do that--, the entropy may change;
that is, the entropy of the \textit{system}~$(1+2)$, when \textit{system}~1 and \textit{system}~2 are isolated, may be different from the sum of the entropies of the parts $S_1 + S_2$.
% , then the entropy of the total \textit{system}~$(1+2)$ is different from the sum of the entropies of the parts $S_1 + S_2$, assuming that the parts are isolated without changing their macroscopic conditions.
If, for isolated systems, $S_{1+2} > S_1 + S_2$, the entropy is said to be \textit{superadditive}, whereas if $S_{1+2} < S_1 + S_2$ the entropy is called \textit{subadditive}~\cite{JStatPhys.52.479.1988.Tsallis, PhysLettA.247.211.1998.Landsberg}.
%
% In general, infinite macroscopic systems are extensive, that is, their equilibrium properties are determined by their extensive parameters (energy, volume, etc) and are independent of the contacts with other systems. The entropy of an extensive system is also extensive, i.e. is a homogeneous function of order one of the extensive parameters.
%
Therefore, the physical properties of mesoscopic systems are influenced (beside their extensive parameters) by their sizes, their contacts with other systems, etc.
% Consequently, their entropy is also not extensive.
For this reasons, the entropy cannot have an universal expression in terms of the extensive parameters, but should be adapted to describe specific types of systems, with specific interactions with the environment.

% The main aim of this paper is to prove that the probability distribution of the closed or open mesoscopic systems corresponding to the nonextensive entropies can be found only by the method based on the principle of maximum entropy or the principle of additivity of the maximal entropy. Moreover, we will show that the method of maximum entropy is a corollary of the method of additivity of the maximal entropy.  

The aim of this paper is to provide a general method, applicable to both, extensive and non-extensive entropies, for the calculation of the equilibrium probability distribution of a system over its microstates.
First, we identify three methods (hypotheses) for the calculation of the  equilibrium probability distributions, which we apply to four well known expressions of for the entropy: the Boltzmann-Gibbs (BG), R\'{e}nyi (R), Havrda-Charv\'{a}t-Daroczy-Tsallis (HCDT), and Landsberg-Vedral (LV) entropies.
Two of these entropies are  extensive (BG and R), whereas the other two are non-extensive (HCDT and LV).
We shall establish the relations between the three methods, to see which of them are physically justified and generally applicable.

The paper is organized as follows.
In Section~\ref{sec_gen_meth} we introduce the methods and  the four entropies BG, R, HCDT, and LV.
In Section~\ref{sec_can_ens} we apply the methods to the  entropies and analyze the results.
In Section~\ref{sec_conclusions} we draw the  conclusions.

\section{General methods} \label{sec_gen_meth}

For the statistical description of the entropy (in both, extensive and non-extensive systems), we need the microstates of the system.
The microstates most commonly used are the eigenstates of the total Hamiltonian of the system.
These microstates will be numbered by $i, j = 0,1, \ldots$ and we denote by $p_i$ the probability to find the system in state $i$.
The entropy is defined as a function of the set of probabilities $\{p_i\}$, so we can write $S \equiv S(\{p_i\})$.
The average of any quantity $A_i$, which is a function of the microstate $i$, is defined as
\begin{equation}
  \langle A \rangle \equiv \sum_i A_i p_i . \label{av_A}
\end{equation}

There are several proposals for the function $S \equiv S(\{p_i\})$.
For extensive systems (but not only), the most commonly used are the Gibbs and Boltzmann entropies~\cite{PhysRev.106.620.1957.Jaynes, PhysRev.108.171.1957.Jaynes, RevModPhys.50.221.1978.Wehrl}.
We do not comment here on which of these two expressions is preferable (see, for example, Refs.~\cite{AmJPhys.33.391.1965.Jaynes, Nat.Phys.10.67.2014.Dunkel, PhysRevE.90.062116.2014.Hilbert, PhysRevE.91.052147.2015.Campisi, AmJPhys.83.163.2015.Frenkel, PhysRevE.92.020103.2015.Swendsen, EPJWebofConf.108.02007.2016.Anghel, PhysicaA.453.24.2016.Swendsen, PhysicaA.467.474.2017.Matty, PhysRevE.95.012125.2017.Abraham}
for a heated discussion), but we shall call it the Boltzmann-Gibbs entropy and write
\begin{equation}
  S^{\rm BG} \equiv - k_B \sum_i p_i \ln p_i. \label{S_Boltzmann}
\end{equation}
The expression~(\ref{S_Boltzmann}) is a particular case--a limit--of a more general additive entropy due to R\'enyi~\cite{ProcFourthBerkeleySymposium.1.547.1061.Renyi, Renyi:book}
\begin{equation}
 S^{\rm R}_q = k_{B}\frac{\ln \left(\sum_{i} p_{i}^{q}\right)}{1-q} . \label{S_Renyi}
\end{equation}
For non-extensive systems, the situation is more complicated, as we explained before.
In this paper %, beside the Boltzmann-Gibbs and R\'enyi entropies,
we shall analyze the Havrda-Charv\'{a}t-Daroczy-Tsallis (HCDT) entropy~\cite{Kybernetika.3.30.1967.Havrda, InfControl.16.36.1970.Daroczy, RevModPhys.50.221.1978.Wehrl, JStatPhys.52.479.1988.Tsallis, PhysicaA.261.534.1998.Tsallis}
\begin{equation}
  S^{\rm HCDT}_q = k_{B} \frac{\sum\limits_{i} p_{i}^{q}-1}{1-q} \label{S_Tsallis}
\end{equation}
and the Landsberg-Vedral entropy~\cite{PhysLettA.247.211.1998.Landsberg}
\begin{equation}
  S^{\rm LV}_q = \frac{k_{B}}{1-q} \left(1-\frac{1}{\sum\limits_{i} p_{i}^{q}}\right) ; \label{S_Landsberg}
\end{equation}
in the expressions (\ref{S_Renyi}), (\ref{S_Tsallis}), and (\ref{S_Landsberg}), $q$ is positive parameter and one can easily check that
$S^{\rm BG}(\{p_i\}) = \lim_{q=1} S^{\rm R}_q(\{p_i\}) = \lim_{q=1} S^{\rm HCTD}_q(\{p_i\}) = \lim_{q=1} S^{\rm LV}_q(\{p_i\})$.

For isolated systems (microcanonical ensemble), the entropies (\ref{S_Boltzmann})-(\ref{S_Landsberg}) are maximized if all accessible microstates are equally probable~\cite{ProcFourthBerkeleySymposium.1.547.1061.Renyi, Renyi:book, JStatPhys.52.479.1988.Tsallis, PhysicaA.261.534.1998.Tsallis, PhysLettA.247.211.1998.Landsberg}.
This means that if we denote by $\Omega \equiv \Omega(E,\bX)$ the number of accessible microstates for fixed external parameters $(E,\bX)$, then $p_i = 1/\Omega = {\rm constant}$ for any $i$.
Therefore, we may write the entropy as a function of $(E,\bX)$, namely $S(E,\bX) \equiv S[\{p_i=1/\Omega(E,\bX)\}]$, where $S$ may take any of the expressions (\ref{S_Boltzmann})-(\ref{S_Landsberg}).
Furthermore, let us assume that \textit{system}~1, with probability distribution $\{p^{(1)}_i\}$, and \textit{system}~2, with probability distribution $\{p^{(2)}_j\}$, are isolated from each-other.
Then, the probability to find \textit{system}~1 in the state $i$ and  \textit{system}~2 in the state $j$ is $p^{(1+2)}_{ij} = p^{(1)}_i p^{(2)}_j$ (statistical independence).
The entropy of the composed \textit{system}~$(1+2)$ is a function of the composed probabilities, $S_{1+2}(\{p^{(1+2)}_{ij}\})$.
If each of these two systems is in microcanonical conditions (i.e. they are isolated from each-other and from the environment), with the external parameters $(E_1,\bX_1)$ (for \textit{system}~1) and $(E_2,\bX_2)$ (for \textit{system}~2), then all the microstates of the composed \textit{system} $(1+2)$ are equiprobable, namely $p^{(1+2)}_{ij} = 1/[\Omega_1(E_1,\bX_1)\Omega_2(E_2,\bX_2)]$, where $\Omega_{1+2} = \Omega_1(E_1,\bX_1)\Omega_2(E_2,\bX_2)$ is the total number of accessible states in the \textit{system}~$(1+2)$.

If systems 1 and 2 interact, so that they can exchange some of their external parameters, then the probabilities $\{p^{(1)}_i\}$ and $\{p^{(2)}_j\}$ are not independent anymore and $\Omega_{1+2}$ is in general different from $\Omega_1(E_1,\bX_1)\Omega_2(E_2,\bX_2)$.
Considering this, we may imagine three general methods to maximize the entropy of a compound system, depending on its physical properties.

The \textit{first method} (M1) is the most commonly used and is generally (but not only) applied to extensive systems.
In this case, the number of states in a system is determined by its external parameters $(E,\bX)$ and does not depend on the contacts with other systems.
For $(E,\bX)$ fixed, all accessible microstates $\Omega(E,\bX)$ are equiprobable and the entropy is $S[\{p_i = 1/\Omega(E,\bX)\}] \equiv S(E,\bX)$.
Let's assume that \textit{system}~1 and \textit{system}~2, of parameters $(E_1,\bX_1)$ and $(E_2,\bX_2)$, respectively, are brought into contact and they can exchange energy.
In this case, the fixed parameters are $\bX_1$, $\bX_2$, and the total energy $E_t \equiv E_1 + E_2$.
% The entropy is determined by the probabilities of all accessible states.
%
% For example, if two such systems are in contact and can exchange energy, and if \textit{system}~1 has the parameters $(E_1, \bX_1)$ and \textit{system}~2 has the parameters $(E_2, \bX_2)$, then $\bX_1$, $\bX_2$, and $E_t \equiv E_1 + E_2$ are constant.
Then, the entropy of the total \textit{system}~$(1+2)$ is calculated from the probabilities of all accessible microstates in the isolated systems 1 and 2, for all possible values of $E_1$ and $E_2$:
\begin{subequations} \label{M1}
\begin{equation}
  S_{1+2, {\rm M1}}(E_t, \bX_1, \bX_2) = S[\{p_{ij}^{(1+2) {\rm M1}}(E_t, \bX_1, \bX_2)\}] , \label{M1_S}
\end{equation}
where the microstate $i$, of \textit{system}~1, corresponds to the external parameters $(E_1, \bX_1)$, the microstate $j$, of \textit{system}~2, corresponds to the external parameters $(E_t - E_1, \bX_2)$, and $E_1$ takes all the physically possible values.
The total number of states in the \textit{system} $(1+2)$ is
\begin{equation}
  \Omega_{1+2, {\rm M1}}(E_t, \bX_1, \bX_2) \equiv \sum_{E_1} \Omega_1(E_1, \bX_1) \Omega_2(E_t-E_1, \bX_2) , \label{M1_O}
\end{equation}
where the summation is taken over all possible values of $E_1$.
Since the \textit{system} $(1+2)$ is isolated, all accessible states are equiprobable and the probabilities of the microstates are
\begin{equation}
  p_{ij}^{(1+2) {\rm M1}}(E_t, \bX_1, \bX_2) = 1/ \Omega_{1+2, {\rm M1}}(E_t, \bX_1, \bX_2) . \label{M1_pij}
\end{equation}
The probabilities~(\ref{M1_pij}) should determine the  value of the entropy of the compound system.

In practice, Eq.~(\ref{M1_pij}) does not represent the  final answer for the equilibrium probability distribution.
If this were the case, then we could not distinguish between the entropies of equilibrium and non-equilibrium configurations, because all would be given by the probabilities~(\ref{M1_pij}).
At this point, an observer comes into play.
If, after merging \textit{system}~1 and \textit{system}~2, the \textit{system}~$(1+2)$ is isolated and  cannot be observed, we do not know anything about the evolution towards equilibrium and Eq.~(\ref{M1_pij}) represents our knowledge.
Nevertheless, in general, we observe that the \textit{system}~$(1+2)$ evolves towards equilibrium and remains there, eventually the extensive parameters of the subsystems having some (small) fluctuations around the equilibrium values.
% 
% during an observation, the energies of the two systems do not vary over the entire accessible range. In general, they have (small) fluctuations around an equilibrium value, which is very close to the most probable value.
We denote the average values by $U_1 \equiv \langle E_1 \rangle$ and $U_2 \equiv \langle E_2 \rangle = E_t - U_1$ and assume that they are the same as the most probable values.
The probability that \textit{system}~1 and \textit{system}~2 have the energies $E_1$ and $E_t - E_1$ is $P(E_1) = \Omega_1(E_1, \bX_1) \Omega_2(E_t-E_1, \bX_2)/ \Omega_{1+2, {\rm M1}}(E_t, \bX_1, \bX_2)$.
Assuming that $P(E_1)$ is a continuous and differentiable function of $E_1$, then $U_1$ and $U_2$ are calculated from the condition of maximum probability,
\begin{eqnarray}
  \frac{d P(E_1)}{dE_1} &\equiv& \frac{\partial [\Omega_1(E_1, \bX_1) \Omega_2(E_t-E_1, \bX_2)]}{\partial E_1} \nonumber \\
  && \times \left[ \Omega_{1+2, {\rm M1}}(E_t, \bX_1, \bX_2) \right]^{-1} = 0 ,
%   &=& \frac{\partial \Omega_1(E_1, \bX_1)}{\partial E_1}  \Omega_2(E_t-E_1, \bX_2) \nonumber \\
%   %
%   && - \Omega_1(E_1, \bX_1) \left. \frac{\partial \Omega_2(E_2, \bX_2)}{\partial E_2} \right|_{E_2 = E_t - E_1} ,
  \label{M1_max_prob}
\end{eqnarray}
which implies
\begin{eqnarray}
  \left. \frac{\partial \ln[\Omega_1(E_1, \bX_1)]}{\partial E_1} \right|_{E_1 = U_1} &=& \left. \frac{\partial \ln[\Omega_2(E_2, \bX_2)]}{\partial E_2} \right|_{E_2 = U_2 = E_t - U_1} \nonumber \\
  &\equiv& 1/ (\kb T_{\rm M1}) \equiv \beta_{\rm M1} . \label{M1_def_T}
\end{eqnarray}
\end{subequations}
Equation~(\ref{M1_def_T}) defines the temperature of the system for the \textit{first method}, $T_{\rm M1}$.
We see also that in deriving Eqs.~(\ref{M1_max_prob}) and~(\ref{M1_def_T}) we assumed that \textit{system}~1 and \textit{system}~2 are not further divided into subsystems and their own evolutions towards equilibrium are not taken into consideration--i.e. they are not observed.
% Equations~(\ref{M1_max_prob}) and~(\ref{M1_def_T}) further imply that the quantity that is maximized in equilibrium is $\ln[\Omega_i(E_1, \bX_1)]$

If the size of one of these systems increases to infinity, it becomes a reservoir and one of the  solutions for this case is always the Gibbs probability distribution in the system that remains finite, no matter what is the definition of the entropy (\ref{S_Boltzmann})-(\ref{S_Landsberg}), as we shall see in Section~(\ref{subsec_M1}).

The \textit{second method} (M2) is due to Jaynes~\cite{PhysRev.106.620.1957.Jaynes}, who applied it to the Boltzmann-Gibbs entropy~(\ref{S_Boltzmann}).
Later, Tsallis applied it to the HCDT entropy~\cite{JStatPhys.52.479.1988.Tsallis}.
So, let's assume that the system under study is in contact with a heat reservoir. Then, its energy is not fixed, but it has--in equilibrium--a well defined average value.
Jaynes proposed that the equilibrium probability distribution is the one that maximizes the entropy, under the constraint of fixed average energy (maximum entropy principle).
Introducing the Lagrange multiplier $1/T_{\rm M2}$, where $T_{\rm M2}$ plays the role of temperature, one finds the equilibrium value of $E$ from the maximization of the (Massieu) function
\begin{subequations} \label{M2}
\begin{eqnarray}
  \Psi \equiv \max_{E} [ S(E,\bX) - E/T_{\rm M2} ] . \label{M2_S}
\end{eqnarray}
Equation~(\ref{M2_S}) implies that
\begin{eqnarray}
  \frac{\partial S(E,\bX)}{\partial E} = \frac{1}{T_{\rm M2} } \label{M2_T}
\end{eqnarray}
\end{subequations}
and may be extended to any type of reservoir~\cite{PhysRev.106.620.1957.Jaynes}.

The \textit{first} and the \textit{second} methods are not equivalent.
They give the same result for Boltzmann-Gibbs entropy~(\ref{S_Boltzmann}), but not for the other three~(\ref{S_Renyi})-(\ref{S_Landsberg}), as we shall see in the next sections.
% The \textit{third method} is a physical justification of the \textit{second method}.

We propose here a \textit{third method} (M3). % which is equivalent to the \textit{second method} for a system in contact with a reservoir.
% The entropies (\ref{S_Boltzmann})-(\ref{S_Landsberg}) are applied to isolated systems, but not when the systems are brought into contact.
Let's suppose again that we have \textit{system}~1 and \textit{system}~2, of parameters $(E_1, \bX_1)$ and $(E_2, \bX_2)$, respectively.
As before, for fixed external parameters, all the accessible states in each of the systems are equiprobable, namely $p_i^{(1)} = 1/\Omega_1(E_1,\bX_1)$ and $p_j^{(2)} = 1/\Omega_2(E_2,\bX_2)$.
These probabilities define the entropies $S_1(\{p_i^{(1)}\}) \equiv S_1(E_1,\bX_1)$ and $S_2(\{p_j^{(2)}\}) \equiv S_2(E_2,\bX_2)$, but also the entropy of the compound \textit{system}~$(1+2)$, $S_{1+2}(\{p_i^{(1)}p_j^{(2)}\})$.
Let's assume now that the systems are brought into contact and that they can exchange energy, so that $E_t \equiv E_1 +  E_2 = {\rm constant}$.
Then, the entropy of the compound \textit{system}~$(1+2)$ is defined as
\begin{subequations} \label{M3}
\begin{equation}
  S_{1+2, {\rm M3}}(E_t, \bX_1, \bX_2) = \max_{E_1} [S_1(E_1,\bX_1) + S_2(E_t-E_1, \bX_2)] . \label{M3_S}
\end{equation}
If $S_1$ and $S_2$ are continuous functions of $E_1$ and $E_2$, respectively, the maximum is obtain when
\begin{equation}
  \frac{\partial S_1(E_1,\bX_1)}{\partial E_1} \equiv \frac{1}{T_{1,{\rm M3}}} =\frac{\partial S_2(E_2,\bX_2)}{\partial E_2} \equiv \frac{1}{T_{2,{\rm M3}}} , \label{M3_T}
\end{equation}
\end{subequations}
which is the Zeroth Principle of Thermodynamics and introduces the notion of temperature for this method.
When one of the systems is a reservoir (i.e. is very large in comparison to the other system), its temperature is fixed and Eq.~(\ref{M3_S}) reduces to Eq.~(\ref{M2_S}), which proves that the \textit{second method} is a corollary of the \textit{third method}.

The \textit{third method} provides also a physical interpretation of the \textit{second method}.
% If systems are extensive, all three methods give the same results.
%
Apparently, the \textit{third method} introduces a degree of arbitrariness in the definition of the entropy, because it differentiate between a compound system and an isolated one.
In an isolated system, the entropy is calculated directly from the probability distribution over the microstates, which is a constant distribution.
In a compound system, e.g. a system formed of two subsystems in thermal contact, the entropy is calculated by the maximization of the sum of the entropies of the two systems.
Therefore, although the compound system is itself isolated from the environment, apparently, its entropy is not calculated from the probability distribution over its microstates. % which should be also equiprobable.
This contradiction is only apparent, since we do not know how the number of microstates is changed when we put in contact two systems which are initially isolated.
If the entropy is properly chosen, e.g. by fitting it with relevant experimental data, then the procedure presented in Eqs.~(\ref{M3}) allows one to estimate the number of microstates accessible in the compound system.
% Without such an empirical solution, one cannot make this estimate.
This also shows that the \textit{third method} is qualitatively different from the  \textit{first method}--where the number of microstates in \textit{system}~$(1+2)$, when \textit{system}~1 and \textit{system}~2 are in contact, is given by~(\ref{M1_O})--and in non-extensive systems leads to different results.
This is the reason for which, as we shall see, the \textit{second method}, which may be derived form the \textit{third method}, gives different results from the \textit{first method}.
Therefore, the  assumption that both M1 and M2 are correct in non-extensive systems~\cite{JStatPhys.52.479.1988.Tsallis, PhysLettA.247.211.1998.Landsberg} may lead to contradictions.

% The contradiction is solved by what we mentioned above, namely that when two systems are brought into contact, the number of accessible microstates changes in a way we cannot predict and therefore describing their thermal properties relies in ones ingeniousness to choose an entropy which fits the phenomenology.
% Once the entropy is properly chosen, the number of accessible microstates in the composed system may be, in principle, calculated by inverting the expression of the total entropy.

In the next section we shall compare the results of the \textit{first} and the \textit{second method}, applied to the entropies (\ref{S_Boltzmann})-(\ref{S_Landsberg}).
We shall show that only the Boltzmann-Gibbs entropy~(\ref{S_Boltzmann}) gives the same result by both methods.
This means that the \textit{third method} (as implied by the \textit{second method}) leads to the standard results, when applied  to macroscopic systems.
The R\'enyi entropy~(\ref{S_Renyi}), even though it is identical to the Boltzmann-Gibbs entropy in isolated systems, gives different results by M1 and M2.

\section{Canonical ensemble probability distribution} \label{sec_can_ens}

Let us calculate the probability distribution over the microstates of a system in contact with a heat reservoir, by employing the \textit{first} and the \textit{second} methods,
%(Sections~\ref{subsec_M1} and~\ref{subsec_M2}, respectively),
to compare the results.

\subsection{Application of the first method} \label{subsec_M1}

Let us assume that \textit{system}~1 is in contact with a heat reservoir $R$.
We denote the parameters of the system by $(E_1, \bX_1)$ and the parameters of the reservoir by $(E_R, \bX_R)$.
The total \textit{system}~$t$, which is composed by the \textit{system}~1 and the reservoir, is isolated and has the energy $E_t = E_1 + E_R$.
The number of microstates in the \textit{system}~$t$ is $\Omega_t(E_t, \bX_1, \bX_R)$, each microstate with the probability $p_{ij}^{(t){\rm M1}} = 1/\Omega_t(E_t, \bX_1, \bX_R)$ as given by Eqs.~(\ref{M1}).
Then, the probability $p_i^{(1) {\rm M1}} (E_1,\bX_1)$ to find the \textit{system}~1 in the microstate $i$, of energy $E_1$, is obtained by summing $p_{ij}^{(t){\rm M1}}$ over all states $j$ of the reservoir which are compatible with $i$.
Since the states $j$ correspond to the energy $E_R = E_t - E_1$, we have
\begin{eqnarray}
  p_i^{(1){\rm M1}} (E_1,\bX_1) &=& p_{ij}^{(t){\rm M1}}(E_t, \bX_1, \bX_R) \Omega_R(E_t - E_1, \bX_R) \nonumber \\
  &=& \frac{\Omega_R(E_t - E_1, \bX_R)}{\Omega_t(E_t, \bX_1, \bX_R)} , \label{pi_M1}
\end{eqnarray}
where $\Omega_R(E_t - E_1, \bX_R)$ is the number of states of energy $E_R = E_t - E_1$ in the reservoir.

We observe in Eq.~(\ref{pi_M1}) that the probability $p_i^{(1){\rm M1}}$ is independent of our choice of entropy~(\ref{S_Boltzmann})-(\ref{S_Landsberg}).
Eventually, our choice of entropy will influence the dependence of $p_i^{(1){\rm M1}}$ on $E_1$, if we assume that the entropy of the reservoir $S_R$ varies linearly with $E_1$:
\begin{eqnarray}
  S_R(E_R, \bX_R) &=& S_R(E_t, \bX_R) - \left. \frac{\partial S_R(E_R, \bX_R)}{\partial E_R} \right|_{E_R = E_t} E_1 \nonumber \\
  &\equiv& S_R(E_t, \bX_R) - \frac{E_1}{T}. \label{SR_M1}
\end{eqnarray}

For Boltzmann-Gibbs entropy~(\ref{S_Boltzmann}) we obtain
\begin{equation*}
  S^{\rm BG}_R = \kb \ln[\Omega_R (E_R, \bX_R)] ,
\end{equation*}
which implies
\begin{subequations} \label{pij_M1}
\begin{equation}
  \Omega_R (E_R, \bX_R) = e^{S_R^{\rm BG}(E_R, \bX_R)/\kb} \equiv e^{\frac{S^{\rm BG}_R(E_t, \bX_R)}{\kb} - \frac{E_1}{\kb T}}. \label{SR_O_B}
\end{equation}
The temperature, in this case, is
\begin{eqnarray}
  \frac{\partial S^{\rm BG}_R(E_R, \bX_R)}{\partial E_R} = \frac{1}{T} \equiv \frac{1}{T_{\rm M1}} \label{SR_M1_B}
\end{eqnarray}
(see Eq.~\ref{M1_def_T}) and combining Eqs.~(\ref{pi_M1}), (\ref{SR_O_B}), and (\ref{SR_M1_B}) we obtain the well-known results
\begin{eqnarray}
  p_i^{(1){\rm BG, M1}} (E_1,\bX_1) = \frac{e^{- E_1/(\kb T_{\rm M1})}}{Z^{\rm M1}_{\rm BG}} \quad {\rm and} \label{pij_M1_B} \\
  Z^{\rm M1}_{\rm BG} \equiv \sum_i e^{- E_1/(\kb T_{\rm M1})} . \nonumber %\label{pij_M1_B_Z}
\end{eqnarray}
\end{subequations}
$Z^{\rm M1}_{\rm BG}$ is a normalization constant--in this case is the partition function--and by $E_1$ we always denote the energy of the state $i$ of the \textit{system}~1, i.e. it is  the  shorthand notation of $E_1(i)$.
% and $E_1$ is the energy of state $i$ in the \textit{system}~1.

Similarly, in the R\'enyi statistics~(\ref{S_Renyi}) one can easily check that
\begin{equation*}
  S^{\rm R}_R = \kb \ln[\Omega_R (E_R, \bX_R)] ,
\end{equation*}
so $\Omega_R^{\rm R}(E_R, \bX_R) = e^{S_R^{\rm R}(E_R, \bX_R)/\kb}$. Repeating the calculations~(\ref{pij_M1}), we obtain
\begin{equation}
  p_i^{(1){\rm R, M1}} (E_1,\bX_1) = \frac{e^{- E_1/(\kb T_{\rm M1})}}{Z^{\rm M1}_{\rm BG}} . \label{pij_M1_R}
\end{equation}

For the HCDT entropy, from Eq.~(\ref{S_Tsallis}) we have
\begin{subequations} \label{HCDT_M1}
\begin{eqnarray}
  S_R^{\rm HCDT} &=& \frac{\kb}{1-q} \left( \Omega_R^{1-q} - 1 \right) \ {\rm and} \label{HCDT_M1_SR} \\
  \Omega_R &=& \left( 1 + \frac{1-q}{\kB} S_R^{\rm HCDT} \right)^{1/(1-q)} \label{HCDT_M1_O} .
\end{eqnarray}
In this case, we can make two hypotheses.
In the \textit{first hypothesis}, we assume that $(\kb T_{\rm M1})^{-1} \equiv \partial [\ln(\Omega_R)] / \partial E_R$ is a constant for any accessible values of $E_R = E_t - E_1$--Abe et al.~\cite{PhysLettA.281.126.2001.Abe} and Toral~\cite{PhysicaA.317.209.2003.Toral} consider that $T_{\rm 1M}$ is the ``physical temperature'', eventually due to its physical interpretation (Eqs.~\ref{M1}) and its identification with the Boltzmann temperature~(Eq.~\ref{SR_M1_B}).
Under this assumption, it follows immediately that
\begin{equation}
  \ln (\Omega_R) = \frac{E_R}{\kb T_{\rm M1}} + C , \label{HCDT_M1_lnO}
\end{equation}
where $C$ is a constant of integration.
Plugging~(\ref{HCDT_M1_lnO}) into Eq.~(\ref{pi_M1}) and using $E_R = E_t - E_1$, we recover the same exponential dependence on energy of the probability distribution,
\begin{equation}
  p_i^{(1){\rm HCDT, M1}} (E_1,\bX_1) = \frac{e^{- E_1/(\kb T_{\rm M1})}}{Z^{\rm M1}_{\rm BG}} , \label{pij_M1_T}
\end{equation}
\end{subequations}
as in Eqs.~(\ref{pij_M1}) and (\ref{pij_M1_R}).

In the \textit{second hypothesis} we employ Eq.~(\ref{SR_M1}), namely, we assume that $T_{R,{\rm HCDT}}^{-1} \equiv \partial S_R^{\rm HCDT} / \partial E_R$ is constant for any accessible $E_R = E_t - E_1$.
From Eq.~(\ref{HCDT_M1_O}) we obtain
\begin{subequations} \label{HCDT_M1_C2}
\begin{eqnarray*}
  \frac{\partial \Omega_R(E_R, \bX_R)}{\partial E_R} &=& \frac{1}{\kb} \frac{\partial S_R^{\rm HCDT}}{\partial E_R} \left( 1 + \frac{1-q}{\kB} S_R^{\rm HCDT} \right)^{q/(1-q)} \\
  &=& \frac{\Omega_R^q(E_R, \bX_R)}{\kb T_{R,{\rm HCDT}}} ,
\end{eqnarray*}
which, after integration, gives
\begin{equation}
  \frac{ \Omega_R^{1-q}(E_t, \bX_R) - \Omega_R^{1-q}(E_R, \bX_R) }{1-q} = \frac{E_t - E_R}{\kb T_{R,{\rm HCDT}}} \label{HCDT_M1_C2_Ogen1}
\end{equation}
which can be written as
\begin{eqnarray}
  && \Omega_R(E_t - E_1, \bX_R) = \Omega_R(E_t, \bX_R) \nonumber \\
  && \times \left[ 1 - \frac{(1-q)E_1}{\kb T_{R,{\rm HCDT}} \Omega_R^{1-q}(E_t, \bX_R)} \right]^{\frac{1}{1-q}} . \label{HCDT_M1_C2_Ogen2}
\end{eqnarray}
Plugging~(\ref{HCDT_M1_C2_Ogen2}) into~(\ref{pi_M1}), we obtain the probability distribution
\begin{eqnarray}
  && {p_i'}^{(1){\rm HCDT, M1}} (E_1,\bX_1) = \frac{1}{Z_{\rm HCDT}^{\rm M1}} \label{pij_M1_T_c2} \\
  && \times \left[ 1 - \frac{(1-q)E_1}{\kb T_{R,{\rm HCDT}} \Omega_R^{1-q}(E_t, \bX_R)} \right]^{\frac{1}{1-q}} , \nonumber
\end{eqnarray}
where $Z_{\rm HCDT}^{\rm M1}$ is a normalization constant and $\Omega_R(E_t, \bX_R)$ is a parameter which has to be determined from experimental observations (the fit of the probability distribution on energy).
When $q \to 1$, $\Omega_R(E_t - E_1, \bX_R)$ and ${p_i'}^{(1){\rm HCDT, M1}} (E_1,\bX_1)$ converge to
\begin{eqnarray}
  \Omega_R(E_t - E_1, \bX_R) &=& \Omega_R(E_t, \bX_R) e^{\frac{- E_1}{\kb T_{{\rm M1}}}} \label{HCDT_M1_C2_Ogen3}
\end{eqnarray}
and
\begin{eqnarray}
  {p_i'}^{(1){\rm HCDT, M1}} (E_1,\bX_1) &=& \frac{ e^{ - E_1 / (\kb T_{{\rm M1}})}}{Z_{\rm BG}^{\rm M1}} \label{HCDT_M1_C2_pp}
\end{eqnarray}
\end{subequations}
which is the expected Boltzmann-Gibbs limit.

For the Landsberg-Vedral entropy, the relation between the entropy and the number of accessible states is
\begin{subequations} \label{LV_M1}
\begin{eqnarray}
  S^{\rm LV}_R = \frac{k_{B}}{1-q} \left(1- \Omega_R^{q-1}\right) \ {\rm and} \label{LV_M1_SR} \\
  \Omega_R = \left(1- \frac{1-q}{\kb} S^{\rm LV}_R\right)^{\frac{1}{q-1}}. \label{LV_M1_O}
\end{eqnarray}
We make again two hypotheses, like for the HCDT entropy.
The \textit{first hypothesis} is that $T_{\rm M1}$ is constant and we obtain again the Gibbs distribution, like in~(\ref{pij_M1_B}), (\ref{pij_M1_R}), and (\ref{pij_M1_T}):
\begin{equation}
  p_i^{(1){\rm LV, M1}} (E_1,\bX_1) = \frac{e^{- E_1/(\kb T_{\rm M1})}}{Z^{\rm M1}_{\rm BG}} . \label{LV_pij_M1_T}
\end{equation}
\end{subequations}

The \textit{second hypothesis} is that $T_{R,{\rm LV}}^{-1} \equiv \partial S_R^{\rm LV} / \partial E_R$ is constant and we obtain
\begin{subequations} \label{LV_M1_C2}
\begin{equation*}
  \frac{\partial \Omega_R(E_R, \bX_R)}{\partial E_R} = \frac{\Omega_R^{2-q}(E_R, \bX_R)}{\kb T_{R,{\rm LV}}} .
\end{equation*}
Applying the same procedure as in Eqs.~(\ref{HCDT_M1_C2}), we obtain
\begin{eqnarray}
  && \Omega_R(E_t - E_1, \bX_R) = \Omega_R(E_t, \bX_R) \nonumber \\
  && \times \left[ 1 - \frac{(q-1)E_1}{\kb T_{R,{\rm LV}} \Omega_R^{q-1}(E_t, \bX_R)} \right]^{\frac{1}{q-1}} . \label{LV_M1_C2_Ogen}
\end{eqnarray}
%
% %
% \begin{equation}
%   \Omega_R(E_t - E_1, \bX_R) = (q-1)^{\frac{1}{q-1}} \left( \frac{E_0 - E_1}{\kb T_{R,{\rm LV}}} \right)^{\frac{1}{q-1}} , \label{LV_M1_C2_Ogen}
% \end{equation}
% %
% where, again, $E_0 \equiv E_t + E_c$ and $E_c$ is a constant of integration.
Equation~(\ref{LV_M1_C2_Ogen}) gives
\begin{eqnarray}
  && {p_i'}^{(1){\rm LV, M1}} (E_1,\bX_1) = \frac{1}{Z_{\rm LV}^{\rm M1}} \label{LV_pij_M1_T_c2} \\
  && \times \left[ 1 - \frac{(q-1)E_1}{\kb T_{R,{\rm LV}} \Omega_R^{q-1}(E_t, \bX_R)} \right]^{\frac{1}{q-1}} , \nonumber
\end{eqnarray}
%
% %
% \begin{equation}
%   {p_i'}^{(1){\rm LV, M1}} (E_1,\bX_1) = \frac{1}{Z_{\rm LV}} \left( \frac{E_0 - E_1}{\kb T_{R,{\rm LV}}} \right)^{\frac{1}{q-1}} , \label{LV_pij_M1_T_c2}
% \end{equation}
% %
where $Z_{\rm LV}^{\rm M1}$ is a normalization constant and $\Omega_R(E_t, \bX_R)$ has to be determined from experimental observations (like in the case of the HCDT entropy).
In the limit $q \to 1$, $\Omega_R(E_t - E_1, \bX_R)$ and ${p_i'}^{(1){\rm LV, M1}} (E_1,\bX_1)$ converge to the Boltzmann-Gibbs expressions,
\begin{eqnarray}
  \Omega_R(E_t - E_1, \bX_R) &=& \Omega_R(E_t, \bX_R) e^{\frac{- E_1}{\kb T_{{\rm M1}}}} \label{LV_M1_C2_Ogen3}
\end{eqnarray}
and
\begin{eqnarray}
  {p_i'}^{(1){\rm LV, M1}} (E_1,\bX_1) &=& \frac{ e^{ - E_1 / (\kb T_{{\rm M1}})}}{Z_{\rm BG}^{\rm M1}} . \label{LV_M1_C2_pp}
\end{eqnarray}
\end{subequations}
%

% \section{Canonical ensemble}\label{sec:2}
% \subsection{Probability distribution function from the maximum entropy principle}

\subsection{Application of the second method} \label{subsec_M2}

In this section we apply the \textit{second method}~\cite{PhysRev.106.620.1957.Jaynes,JStatPhys.52.479.1988.Tsallis,PhysicaA.261.534.1998.Tsallis} to calculate the probability distribution of \textit{system}~1 in contact with a heat reservoir $R$ of temperature $T$ (the canonical ensemble). \textit{System}~1 is described by one of the entropies~(\ref{S_Boltzmann})-(\ref{S_Landsberg}).
In order to do this, we solve~Eq.~(\ref{M2_S}), with the constraint
\begin{equation}\label{5}
  \sum\limits_{i}  p_{i} = 1.
\end{equation}
The Boltzmann-Gibbs entropy~(\ref{S_Boltzmann}) leads to
\begin{subequations} \label{M2_B}
\begin{eqnarray}
  p^{(1) {\rm B, M2}}_{i} &=& \frac{1}{Z^{\rm M2}_{\rm BG}} e^{-E_{i}/(\kb T)}, \ {\rm where} \label{10} \\
  Z^{\rm M2}_{\rm BG} &=& \sum\limits_{i} e^{-E_{i}/(\kb T)} \equiv Z^{\rm M1}_{\rm BG}. \label{11}
%   \left\langle  A \right\rangle &=& \frac{1}{Z_{C}} \sum\limits_{i} A_{i}  e^{-\frac{E_{i}}{T}}. \label{12}
\end{eqnarray}
\end{subequations}
We observe that the probability distribution and the partition function~(\ref{M2_B}) are identical to the ones obtained by the \textit{first method}~(\ref{pij_M1}), as expected.

For the R\'{e}nyi entropy~(\ref{S_Renyi}), we obtain by the \textit{second method}~\cite{PhysLettA.340.375.2005.Parvan, PhysLettA.374.1951.2010.Parvan}
\begin{subequations} \label{M2_R}
\begin{eqnarray}
  p^{(1) {\rm R, M2}}_{i} &=& \frac{1}{Z^{\rm M2}_{\rm R}}\left[1+(q-1)\frac{\Lambda_{\rm R}-E_{i}}{\kb T}\right]^{\frac{1}{q-1}}  \label{13} \\
  {\rm and} \quad Z^{\rm M2}_{\rm R} &=& \sum\limits_{i}  \left[1+(q-1)\frac{\Lambda_{\rm R} - E_{i}}{\kb T}\right]^{\frac{1}{q-1}}, \label{14}
%   \left\langle  A \right\rangle &=& \frac{1}{Z_{C}} \sum\limits_{i} A_{i}  \left[1+(q-1)\frac{\Lambda-E_{i}}{T}\right]^{\frac{1}{q-1}} \label{15}
\end{eqnarray}
where
\begin{equation}
  \left( Z^{\rm M2}_{\rm R} \right)^{q-1} = \frac{q}{\chi}
  \quad {\rm and} \quad
  \chi\equiv \sum\limits_{i} \left( p^{(1) {\rm R, M2}}_{i} \right)^{q} . \label{16}
\end{equation}
%and $\Lambda_{\rm R}=\lambda- \kb T/(q-1)$.
%
% where $\Lambda_{\rm R}=\lambda- \kb T/(q-1)$.
%\dragos{Nu inteleg de ce mai avem nevoie si de relatia $\Lambda=\lambda-T/(q-1)$. Putem sa o stergem?}
From the Eqs.~(\ref{M2_R}) above we obtain a selfconsistent equation for $\Lambda_{\rm R}$,
% Substituting eq.~(\ref{13}) into $\chi$ and using eqs.~(\ref{14}), (\ref{16}), we find an equation for $\Lambda$ as
\begin{eqnarray}\label{17}
  && \sum\limits_{i}  \left[1+(q-1)\frac{\Lambda_{\rm R}-E_{i}}{\kb T}\right]^{\frac{1}{q-1}}    \nonumber \\
  &&  = \frac{1}{q} \sum\limits_{i} \left[1+(q-1)\frac{\Lambda_{\rm R}-E_{i}}{\kb T}\right]^{\frac{q}{q-1}} .
\end{eqnarray}
\end{subequations}
We see that the distribution~(\ref{17}), which is derived from the condition~(\ref{M3_T}), is different from the distribution~(\ref{pij_M1_R}), which is derived from the condition~(\ref{M1_def_T}).

For the HCDT entropy~(\ref{S_Tsallis}) one obtains~\cite{JStatPhys.52.479.1988.Tsallis, PhysicaA.261.534.1998.Tsallis, PhysLettA.226.257.1997.Plastino, PhysLettA360.26.2006.Parvan}
\begin{subequations} \label{M2_HCDT_1}
\begin{eqnarray}
  p^{(1) {\rm HCDT, M2}}_{i} &=& \left[1+\frac{q-1}{q}\frac{\Lambda_{\rm HCDT}-E_{i}}{\kb T}\right]^{\frac{1}{q-1}}, \label{18}
\end{eqnarray}
where $\Lambda_{\rm HCDT}$ satisfies the equation
\begin{equation}
   \sum\limits_{i}  \left[1+\frac{q-1}{q}\frac{\Lambda_{\rm HCDT} -E_{i}}{\kb T}\right]^{\frac{1}{q-1}} = 1 . \label{20}
%   \left\langle  A \right\rangle &=&  \sum\limits_{i} A_{i}  \left[1+\frac{q-1}{q}\frac{\Lambda-E_{i}}{T}\right]^{\frac{1}{q-1}}, \label{19}
\end{equation}
\end{subequations}
%
%where $\Lambda=\lambda- \kb Tq/(q-1)$ \cred{(iarasi avem nevoie de $\lambda$?)}.

In the Tsallis statistics, one can also define generalized expectation values~\cite{JStatPhys.52.479.1988.Tsallis, PhysicaA.261.534.1998.Tsallis, PhysLettA.226.257.1997.Plastino, JPhysA.24.L69.1991.Curado, JPhysA.25.1019.Curado} by
\begin{equation}
  \langle  A \rangle_q =\sum_{i} A_{i}  p_{i}^{q}.
\end{equation}
In this case, the probability distribution function and ensemble averages are~\cite{PhysicaA.261.534.1998.Tsallis,PhysRevLett.85.4691.Gudima}
\begin{subequations} \label{M2_HCDT}
\begin{eqnarray}
  && p^{(1) {\rm HCDT, M2}}_{i,q} = \frac{1}{Z^{\rm M2}_{\rm HCDT}}\left[1-(1-q)\frac{E_{i}}{\kb T}\right]^{\frac{1}{1-q}}, \label{21} \\
  && Z^{\rm M2}_{\rm HCDT} = \sum\limits_{i}  \left[1-(1-q)\frac{E_{i}}{\kb T}\right]^{\frac{1}{1-q}}. \label{22}
  %
%   && \langle  A \rangle_q = \frac{1}{\left( Z^{\rm M2}_{\rm HCDT} \right)^{q}} \sum\limits_{i} A_{i}  \left[1-(1-q)\frac{E_{i}}{\kb T}\right]^{\frac{q}{1-q}}. \label{23}
\end{eqnarray}
\end{subequations}

We observe that the  energy dependence of $ p^{(1) {\rm HCDT, M2}}_{i,q}$ (Eq.~\ref{21}) recovers the energy dependence of ${p_i'}^{(1){\rm HCDT, M1}}$ (Ec.~\ref{pij_M1_T_c2}) if $T\equiv T_{R,{\rm HCDT}} \Omega_R^{1-q}(E_t, \bX_R)$ and $Z^{\rm M2}_{\rm HCDT}\equiv Z^{\rm M1}_{\rm HCDT}$. These equations are similar in form, but the expectation values are different.

Applying the \textit{second method} to the Landsberg-Vedral entropy~(\ref{S_Landsberg}), we obtain~\cite{arXiv171109354.Parvan}
\begin{subequations} \label{M2_LV}
\begin{eqnarray}
  p^{(1) {\rm LV, M2}}_{i} &=& \frac{1}{Z^{\rm M2}_{LV}}\left[1+(q-1)\frac{\Lambda_{\rm LV} - E_{i}}{\kb T}\right]^{\frac{1}{q-1}}, \label{24} \\
  Z^{\rm M2}_{LV} &=& \sum\limits_{i}  \left[1+(q-1)\frac{\Lambda_{\rm LV} - E_{i}}{\kb T}\right]^{\frac{1}{q-1}}, \label{25}
%   \left\langle  A \right\rangle &=& \frac{1}{Z_{C}} \sum\limits_{i} A_{i}  \left[1+(q-1)\frac{\Lambda-E_{i}}{T}\right]^{\frac{1}{q-1}} \label{26}
\end{eqnarray}
where
\begin{equation}\label{27}
  \left( Z^{\rm M2}_{LV} \right)^{q-1} = \frac{q}{\chi^{2}} \quad {\rm and} \quad \chi\equiv \sum\limits_{i} \left(p^{(1) {\rm LV, M2}}_{i}\right)^{q}.
\end{equation}
From Eqs.~(\ref{M2_LV}) we find a self-consistent equation for $\Lambda_{\rm LV}$,
\begin{eqnarray}\label{28}
  && \sum\limits_{i}  \left[1+(q-1)\frac{\Lambda_{\rm LV} - E_{i}}{\kb T}\right]^{\frac{1}{q-1}}    \nonumber \\
  &&  = \left\{q^{-1/2}\sum\limits_{i} \left[1+(q-1)\frac{\Lambda_{\rm LV} - E_{i}}{\kb T}\right]^{\frac{q}{q-1}} \right\}^{\frac{2}{q+1}} .
\end{eqnarray}
\end{subequations}

We observe that the  energy dependence of $p^{(1) {\rm LV, M2}}_{i}$ (Eq.~\ref{24}) resembles the energy dependence of ${p_i'}^{(1){\rm LV, M1}}$ (Ec.~\ref{LV_pij_M1_T_c2}).
To better understand this resemblance, we define
\begin{subequations} \label{M1_M2_LV}
\begin{equation}
  \omega^{q-1} \equiv 1+(q-1)\frac{\Lambda_{\rm LV}}{\kb T} , \label{M1_M2_omega}
\end{equation}
which, if plugged into~(\ref{24}), leads to
\begin{eqnarray}
  p^{(1) {\rm LV, M2}}_{i} &=& \frac{\omega}{Z^{\rm M2}_{LV}}\left[1 - (q-1)\frac{E_{i}}{\kb T \omega^{q-1}}\right]^{\frac{1}{q-1}} . \label{M1_M2_p1}
\end{eqnarray}
\end{subequations}
Equations (\ref{LV_pij_M1_T_c2}) and (\ref{M1_M2_p1}) are similar in form, but since the quantities $\Omega_R^{q-1}(E_t, \bX_R)$ and $\omega$ are different (the  condition~(\ref{28}) imposed on $\Lambda_{\rm LV}$ and therefore on $\omega$ is not necessary for $\Omega_R(E_t, \bX_R)$), the probability distributions are  also different, as we anticipated in Section~\ref{sec_intro}.
Furthermore, we can also see from Eqs.~(\ref{HCDT_M1_SR}) and (\ref{LV_M1_SR}) that by the \textit{first method} the relation
\begin{eqnarray}
  S_{1+2, {\rm M1}}(E_t, \bX_1, \bX_2) &=& S_{1}(E_1, \bX_1) + S_{2}(E_2, \bX_2) \nonumber \\
  && + \gamma S_{1}(E_1, \bX_1) S_{2}(E_2, \bX_2)
  \label{rel_fin}
\end{eqnarray}
is always satisfied--where $\gamma = (1 - q)/\kb$ for the HCDT entropy and $\gamma = - (1 - q)/\kb$ for the LV entropy~\cite{JStatPhys.52.479.1988.Tsallis, PhysLettA.247.211.1998.Landsberg}--and this is in contradiction with Eq.~(\ref{M3_S}) of M3.

\section{Conclusions} \label{sec_conclusions}

%In general, infinite macroscopic systems are extensive, that is, their equilibrium properties are determined by their extensive parameters (energy, volume, etc) and are independent of the contacts with other systems. The entropy of an extensive system is also extensive, i.e. is a homogeneous function of order one of the extensive parameters.
%
%Mesoscopic systems are non-extensive--their sizes, together with the contacts with other systems influence their physical properties, beside their extensive parameters.
%Consequently, their entropy is not extensive.
%For these reasons, the entropy cannot have an universal expression in terms of the extensive parameters, but should be adapted to describe specific types of systems, with specific interactions with the environment.

In this paper we analyzed the extensive Boltzmann-Gibbs (Eq.~\ref{S_Boltzmann}) and R\'enyi (Eq.~\ref{S_Renyi}) entropies, together with the non-extensive Havrda-Charv\'{a}t-Daroczy-Tsallis (Eq.~\ref{S_Tsallis}) and Landsberg-Vedral entropies (Eq.~\ref{S_Landsberg}).
We calculated the canonical distributions by two methods.
In the \textit{first method} (M1) we assumed statistical independence of the systems, that is, if \textit{system}~1 and \textit{system}~2, of extensive parameters $(E_1, \bX_1)$ and $(E_2, \bX_2)$, respectively, are in contact and in equilibrium, then the number of accessible microstates in the total \textit{system}~$(1+2)$ is $\Omega_{1+2} = \Omega_1(E_1.\bX_1) \Omega_2(E_2.\bX_2)$, where $\Omega_1(E_1, \bX_1)$ and $\Omega_2(E_2.\bX_2)$ are the number of states in \textit{system}~1 and \textit{system}~2, respectively.

In the \textit{second method} (M2) we assumed that the canonical equilibrium probability distribution over the microstates is the one that maximizes the entropic thermodynamic potential, that is, maximizes the entropy under the  restriction that the average energy is fixed.
We have shown that the two methods are not equivalent and they give the same result only for the Boltzmann-Gibbs statistics.

We have also shown that the \textit{second method} is related to the principle of additivity of entropy (called the  \textit{third method} in the text).
We proposed that the entropy should be defined in general (i.e. for both, extensive and  non-extensive systems) such that, when \textit{system}~1, of parameters $(E_1, \bX_1)$, and \textit{system}~2, of parameters $(E_2, \bX_2)$, are put into thermal contact, the entropy of the composed \textit{system}~$(1+2)$ becomes $S_{1+2}(E_t, \bX_1, \bX_2) \equiv \max_{E'_1} [S_1(E'_1, \bX_1) + S_2(E_t-E'_1, \bX_2)]$, where $E_t \equiv E_1 + E_2$.
The \textit{second method} is  a corollary of the \textit{third method}--it is the \textit{third method}, when one of the systems is a reservoir--and these two are different from the \textit{first  method}, which is widely used in macroscopic thermodynamics.
%
% The \textit{second method} is  a corollary of the \textit{third method}, i.e. of the additivity of the entropy, which is, furthermore, not equivalent to the principle of statistical independence (\textit{first  method}), widely used in macroscopic thermodynamics.
%
Therefore we conclude that the principle of additivity of entropy provides a physical foundation for both, extensive and non-extensive statistics, for the method of maximization of the entropic thermodynamic potential, and is of general applicability in both, macroscopic and mesoscopic systems.
On the other hand, it overrules the principle of statistical independence of the probability distributions over the microstates, which is applicable only for isolated systems and is just a convenient approximation in macroscopic systems which are  in contact.

{\bf Acknowledgments:} This work was supported in part by Romania-JINR collaboration projects. DVA was supported by the ANCS project PN18090101/2018.

% \bibliography{general}
% \bibliography{/media/sf_Share/Dropbox/general}
\bibliography{/home/dragos/Dropbox/general}
\bibliographystyle{unsrt}

\end{document}